\documentstyle[art10,hip-100a,epsfig]{article}
\begin{document}

\begin{center}
{\bf INSTITUT~F\"{U}R~KERNPHYSIK,~UNIVERSIT\"{A}T~FRANKFURT}\\
D-60486 Frankfurt, August--Euler--Strasse 6, Germany
\end{center}

\hfill IKF--HENPG/4--96
\vspace{2cm}

\begin{center}
{\Large {\bf
Strangeness Enhancement in Sulphur-Nucleus Collisions at 200 GeV/N
}}
\end{center}


\vspace{2cm}
\begin{center}
J\"urgen Eschke 

\vspace{0.5cm}
\noindent

Gesellschaft f\"ur Schwerionenforschung, Darmstadt, Germany\\
E--mail address: J.Eschke@gsi.de

\vspace{5cm}
\noindent
{\it Talk given at STRANGENESS'96, Budapest, Hungary, May 15--17 1996} \\
{\it to be published in the special issue of Heavy Ion Physics}

\end{center}

\vfill

\newpage
\cimo \setcounter{page}{1} \thispagestyle{empty} \hskip -15pt  \noindent
{\Large\bf
%
Strangeness Enhancement in 
Sulphur-Nucleus\\
Collisions at 200 GeV/N
%
}\\[3mm] 
\def\rightmark{Strangeness Enhancement in Sulphur-Nucleus Collisions}\def\leftmark{J. Eschke}
\hspace*{6.327mm}\begin{minipage}[t]{12.0213cm}{\large\lineskip .75em
J. Eschke$^1$ and the {\bf NA35 Collaboration} 
}\\[2.812mm] 
\hspace*{-8pt}$^1$ GSI Darmstadt,
Planckstr. 1, 64291 Darmstadt, Germany,\\
E-mail: J.Eschke@gsi.de\\[0.2ex]
\\[4.218mm]{\it
Received nn Month Year (to be given by the editors)
}\\[5.624mm]\noindent
{\bf Abstract.} 
The NA35 experiment used several independent methods to determine the strange
particle production in p+S and S+A collisions.
The different techniques show consistent results.
Strangeness conservation in full phase space is used as an additional check of the consistency
of the data.
 On the base of the analysis in full phase space
it could be shown that strangeness conservation is fullfilled.
The NA35 K$^0_S$ in S+S and S+Ag are consistent with the NA44 results for K$^+$ and K$^-$.
The results of the NA36 collaboration for S+Pb collisions were extrapolated to full phase space.
The comparison with the NA35 results shows more than two times lower yields.
The ratio of $\Lambda$ to $\overline{\Lambda}$ at midrapidity of NA36 is inconsistent
with the high  baryon density determind by NA35.
The strange particle production is compared to the abundance of
non strange particles, especially nega\-tive\-ly charged pions which are measured
in full phase space in the same experiment. 
A clear enhanced strange hadron production relative to $\pi^-$ is observed
in S+Ag collisions compared to p+S reactions at the same energy.
The K$^0_S$ multiplicity in full phase space per negative hadron (h$^-$)
in S+S, S+Ag and Pb+Pb is enhanced by about a factor 1.6 compared to N+N and p+S collisions.
The NA36 result for the K$^0_S$ multiplicity per h$^-$ in S+Pb is below the N+N value.
\end{minipage}
\section{Introduction}
The enhanced production of strange hadrons in nucleus-nucleus collisions at
ultra\-relativistic energies was predicted to be a signal of the formation
of a QGP \cite{bib11}.
The experiment NA35 performed a systematic study of the production of
$\Lambda$, $\overline{\Lambda}$, K$^{0}_S$, K$^{+}$ and K$^{-}$ in
p+A and S+A reactions at 200 GeV per nucleon \cite{bib1}.
Charged particles over a wide rapidity range are recorded
by a streamer chamber and a time projection chamber.
Different experimental setups and analysis technique for
the measurement of strange hadrons were used to determine
strange particle yields in
various systems and in different phase space regions.\\
%
%
\hbox to\columnwidth{\hfil 0231-4428/95/ \$ 5.00}
\hbox to\columnwidth{\hfil
\copyright 1996 Akad\'emiai Kiad\'o, Budapest}
\noindent
\section{Experimental procedure}
The schematic set--up of the experiment NA35 is shown in Fig.~1.
The charged particles are detected in a large volume
tracking chamber, the Streamer Chamber  placed inside a 1.5 Tesla
dipole magnet (VTM).
The data were collected
in two different target -- Streamer Chamber configurations.
In the 'standard' configuration the nuclear targets
were placed about 10 cm upstream of
the entry window of the Streamer Chamber.
fig.~1 also shows a streamer chamber picture of a central S+Ag collision.
The rapidity--transverse momentum ($y$--$p_T$)
acceptance of this configuration covers
 the $y$--region  below midrapidity
 ($y_{cm} = 3$) and $p_T > 0.3$ GeV/c for
$\Lambda$ and
 $K^0_S$ and $p_T < 1 $ GeV/c for
 charged kaons.
In the 'sweeper' configuration an additional
 1.8 Tesla sweeper magnet (SM) was installed in front of the VTM
magnet, and the target was
placed 361 cm upstream of the Streamer Chamber.
These modifications lead to  track densities low enough
for an efficient analysis in the entire volume of the chamber.
The acceptance for neutral strange particles in this configuration is limited
to the forward rapidity region (y$>$3) \cite{bib1}. 
\subsection{Analysis techniques}
The strange particles were detected by reconstructing their charged decay
vertices:
$\Lambda \rightarrow p + \pi^-$, $\overline{\Lambda} \rightarrow
\overline{p} + \pi^+$, $K^0_S \rightarrow \pi^+ + \pi^-$,
$K^+ \rightarrow (\mu^+ + \nu_{\mu}, \pi^+ + \pi^0, \pi^+ + \pi^+ + \pi^-)$
and
$K^- \rightarrow (\mu^- + \overline{\nu}_{\mu}, \pi^- + \pi^0,
\pi^- + \pi^- + \pi^+)$.
Two  procedures were used leading to different event samples:
\begin{itemize}
\item
a 'visual' method in which a human operator detected
 and measured the tracks associated with a
 strange particle decay \cite{bib10} and
\item
an 'automatic' method in which digitized Streamer Chamber
images 
were analyzed by a special software tracking chain \cite{bib6}.
\end{itemize}
In the 'visual' method of strange particle analysis the visibility
of  the decay vertex was required. In the 'computerized' method
of  the $\Lambda$, $\overline{\Lambda}$ and  $K^0_S$
analysis the decay point was reconstructed using combinations
of all positive and negative tracks measured in the same
event.
In order to reduce the combinatorial background produced
by tracks originating from the interaction vertex it was necessary
to apply cuts on the  distance of the point of closest
approach of the track
to the interaction vertex
and on the distance between the decay  and  interaction vertices.
 The resulting
invariant mass distributions of the two decay products
for central $^{32}$S+Ag
collisions are plotted in Fig. 2 for the $\Lambda$, $\overline{\Lambda}$
and  $K^0_S$ decay hypothesis.
The combinatorial
background is reduced by these pure geometrical cuts below the 10\%
level.

For further background reduction the candidates for strange particle decays
were fitted using the least--squares method and
tested against the $\Lambda$, $\overline{\Lambda}$ and  $K^0_S$
decay hypotheses using Pearson's $\chi^2$ test.
The cuts on the  $\chi^2$ of
the kinematical fit
and on the angle between the momenta of the decay product and the strange
particle (measured in the decay c.m. system) lead to a suppression
of the background to less than 3\%.
Corrections evaluated by a Monte Carlo simulation of the real data
were used to obtain the final results.
The  $y$--$p_T$ acceptances obtained for the 'automatic'
method are wider than the acceptances for the 'visual'
procedure.
\section{Experimental results}
The rapidity
distributions of $\Lambda$, $\overline{\Lambda}$ and $K^0_S$
from the 'visual' and
'automatic' methods are compared in Fig.~3.
The figures show the comparison of the 'visual' and 'automatic' 
analysis method for S+Ag collisions in the target hemisphere
(y$<$3) and the results of the 'visual' analysis for S+Au reactions
in the projectile hemisphere (y$>$3) obtained in the 'sweeper'
configuration of the experiment.
The results of the different methods are consistent within the
statistical and systematical errors.
The NA35 data for central S+Ag collisions in the target hemisphere
(y$<$3) have been evaluated by averaging the results of the 'visual'
and 'automatic' analysis technique. The distributions for S+Ag
in the projectile hemisphere (y$>$3) have been obtained 
by interpolation between S+S and S+Au data.
The measurement of K$^+$ and K$^-$ \cite{bib3} was used to check the 
consistency between different methods.
The K$^0_S$ yield has to agree with the average of
K$^+$ and K$^-$  (~0.5$\cdot$(K$^+$+K$^-$)~) for an isospin zero system.   
Fig.~4 demonstrates that this is also fullfilled for
the S+Ag system, which has a relatively small charge asymmetry.
The agreement between K$^0_S$ and 0.5$\cdot$(K$^+$+K$^-$) is shown
in a transverse mass distribution and a rapidity distribution
in S+Ag collisions.

The determination of the kaon and hyperon yields in full phase space 
allows us to test whether strangeness is conserved. 
It has been shown on the base of the NA35 results in S+S collisions  \cite{bib3}
that the excess of kaons over antikaons \mbox{(~$<K>_{net}$~$=$~2$\cdot$($<K^+>$-$<K^->$)~)}
 matches the excess of hyperons over antihyperons
\mbox{(~$<Y>_{net}$~$=$~1.6$\cdot$($<\Lambda + \Sigma^0>$-$<\overline{\Lambda} + \overline{\Sigma^0}>$)~)}.
The factor 2 accounts for K$^0$ and $\overline{K^0}$, the scaling factor 1.6 is taken from \cite{bib4}
and accounts for unobserved hyperons.
\subsection{Comparison with NA36 and NA44}
In Fig.~5 the results of NA35 are compared to NA44 and NA36 data for asymmetric systems.
NA44  uses a spectrometer to measures charged particles yields.
Its acceptance is very narrow around rapidity y=3 and reaches down to zero transverse momentum.
The figure shows the agreement of the average   
of K$^+$ and K$^-$  (~0.5$\cdot$(K$^+$+K$^-$)~) of NA44 in S+Pb collisions \cite{bib5} with the NA35
K$^0_S$ yield in S+Ag at midrapidity.
The kaon data of NA35 and NA44 are also consistent for the S+S system \cite{bib5}.

The TPC-experiment NA36 measures 
$\Lambda$, $\overline{\Lambda}$ and K$^0_S$ in S+Pb collisions in a limited $y-p_T$ acceptance.
In order to compare the NA36 data to NA35 results the NA36 distributions
were extrapolated to the full 
transverse momentum range.
This extrapolation was done assuming an exponential behaviour in the transverse mass m$_T$   
down to p$_T$=0 (m$_T$=m$_0$) as demonstrated in Fig.~4.
In Fig.~5 the NA35 result for $\Lambda$ in central S+Ag reactions
is compared to the $\Lambda$ production determined by NA36
in S+Pb collision at 200~GeV per nucleon as presented at Quark Matter 95 \cite{bib2}.

In this reference the NA35 $\Lambda$ rapidity distribution was scaled down,
in order to compare to the NA36 $\Lambda$ distribution in the transverse
momentum range \mbox{$0.4<p_T<1.8$~GeV/c}.
The scaling factor used by NA36 was 0.76.
Also the sligthly softer trigger
of NA36 was taken into account. For the comparison in ref.\cite{bib2} the NA35 $\Lambda$ yield
was scaled down by the ratio of the negative particles \mbox{(~n$^{S+Pb}_{NA36}$$/$n$^{S+Ag}_{NA35}$=0.85~)}
in the two experiments. 
The inverse of these two scaling factors determined by NA36 was used to scale up the
NA36 $\Lambda$ to the full transverse momentum range and correct for the different trigger
(see Fig.~5).

For the comparison with the
$\overline{\Lambda}$ and K$^0_S$ of NA36 in S+Pb collisions
the data from Ref.~\cite{bib8} (Quark Matter 93) were used since
these results were not updated in ref.\cite{bib2}.
The NA36 results for $\Lambda$ between QM93 and QM95
increased only by 16$\%$.
In order to extrapolate the NA36 $\overline{\Lambda}$ and K$^0_S$ from the transverse
momentum range $0.6<p_T<1.6$~GeV/c to the full p$_T$ range we calcultate the extrapolation factor as:
\begin{equation}
w = \frac
{\int^{\infty}_{0} p_T \cdot e^{-m_T/T} \cdot dp_T}
{\int^{1.6}_{0.6} p_T \cdot e^{-m_T/T} \cdot dp_T}.
\end{equation}
where T is the inverse slope of the transverse mass spectrum.

The extrapolation factors were calculated for $\overline{\Lambda}$
and K$^0_S$ using an inverse slope parameter T of 220$\pm$15~MeV.
This T parameter is similar to the values NA35 determined
for strange particles in central S+Ag collisions \cite{bib1}.  
The extrapolation factor $w$ is 1.8$\pm$0.05 for
$\overline{\Lambda}$ and 2.58$\pm$0.2 for K$^0_S$.
The uncertainty for the scaling factors is determined by the error
of the T parameter.
Also the correction for the different trigger was applied.
Fig.~5 shows that the extrapolated NA36 results are more than a factor
2 lower than the NA35 data.
\section{Discussion}
The rapidity distribution
of the 'net protons' ($p-\overline{p}$) for S+S, S+Ag and S+Au \cite{bib7} shown in Fig.~6 
demonstrates that a larger fraction of the participating baryons are stopped in nuclear collisions
at 200~GeV per nucleon and shifted towards midrapidity.
The distribution of
the 'net' $\Lambda$ of NA35 ($\Lambda$-$\overline{\Lambda}$) 
has a similar shape as seen in the 'net protons', because a large
fraction of the momentum of a produced $\Lambda$ is determined by the leading diquark of the proton.
The $\Lambda$-$\overline{\Lambda}$ distribution of NA36 approaches zero at midrapidity, which
is inconsistent with the measured 'net' baryon density.

In Fig.~8 the S+Ag data are compared with particle yields in
	p+S collision at the same energy,
	which were scaled with the ratio of the negative hadron
	multiplicity in the two systems. A clear enhancement of
	about a factor 2 at midrapidity can be seen.
The enhancement is also seen in Fig.~7
 for the K$^0_S$ multiplicity in full phase space per negative hadron (h$^-$)
	for N+N, p+S, S+S, S+Ag and Pb+Pb.
The K$^0_S$ and h$^-$ multiplicities are listed in ref.~\cite{bib1}.
	The Pb+Pb data are preliminary results of experiment NA49 \cite{bib9}.
	The ratio in Nucleus-Nucleus is enhanced by about a factor 1.6 compared to N+N and p+S collisions.
In order to compare to the NA36 result for K$^0_S$ a rough extrapolation to full phase space
was done.
Therefore a similar shape of the NA36 K$^0_S$ rapidity distribution as measured by NA35 was assumed.
This extrapolation has a relativ large uncertainty of about 20$\%$.
The evaluated 4$\pi$ multiplicity for K$^0_S$ is 7$\pm$1.5. The negative particle
multiplicity of NA36 in S+Pb collisions of $\approx$144 was taken from Ref.~\cite{bib2}.
	The NA36 result for the K$^0_S$ multiplicity per negativ hadron in S+Pb is below the N+N value.
\section{Conclusions}
A clear enhanced strange hadron production relative to $\pi^-$ is observed
in S+A collisions compared to p+p and p+A reactions at the same energy.
This strong enhancement points at
a fast strangeness production in the dense reaction zone.
The strange particle abundances indicate a fast approach to chemical equilibrium which is
difficult to explain on the base of pure hadronic interactions.\\

\vfill\eject 
\begin{figure}
        \epsfig{figure=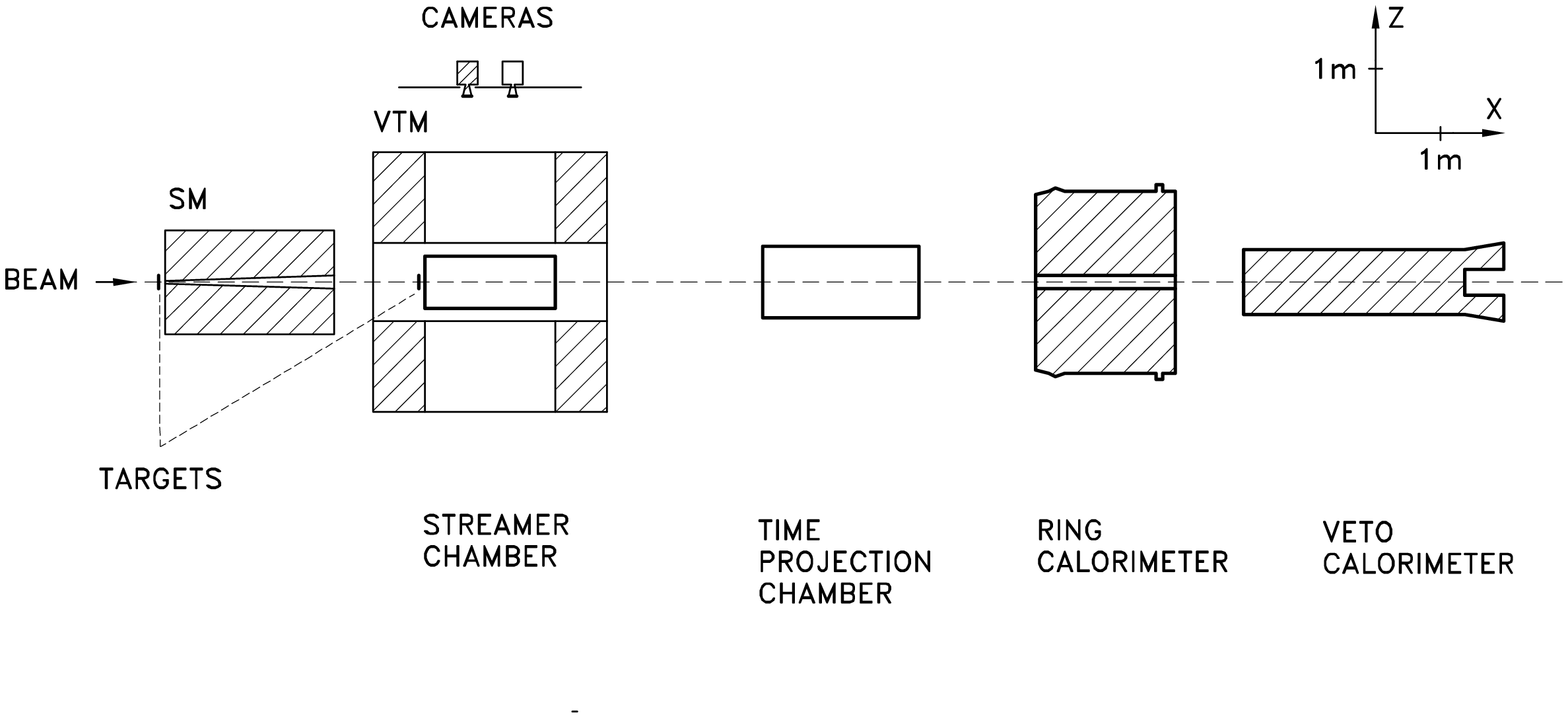,width=12cm,height=4cm,angle=90}
	\vspace{1cm}
        \epsfig{figure=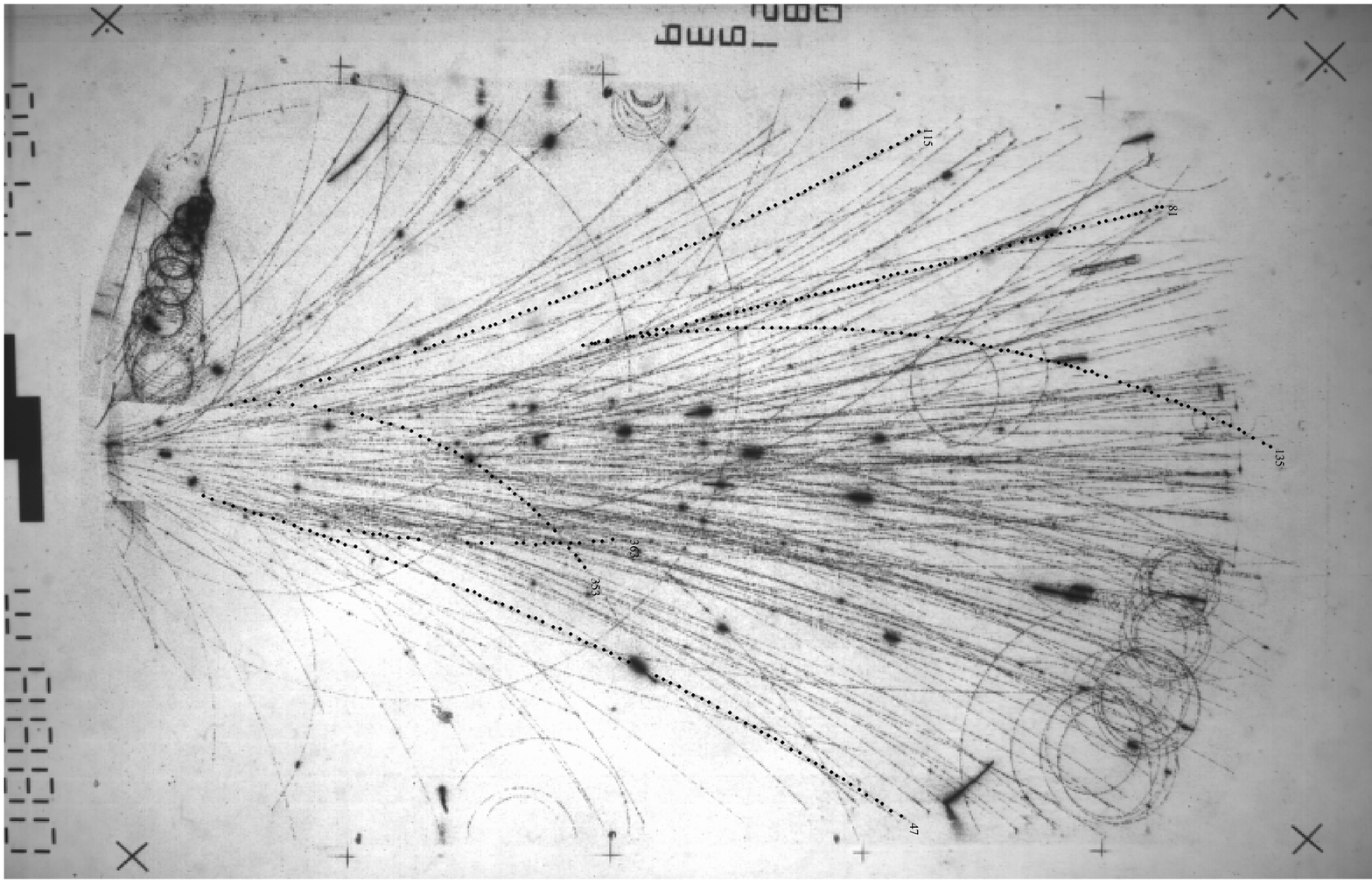,width=12cm,angle=-90}
        \caption[xx]{
Schematic side view  of the NA35 experiment.
The magnetic field in the sweeper magnet (SM) and in the
VTM magnet is parallel to the z axis. The Streamer Chamber is
placed in the center of the VTM magnet. The forward energy deposited
in the Veto Calorimeter was used to select central collisions.
The lower part of the figure shows a streamer chamber picture of a central S+Ag collision.
Tracks belonging to decays of identified neutral strange particles a marked with black dots.}
\end{figure}
\newpage
\begin{figure}
        \epsfig{figure=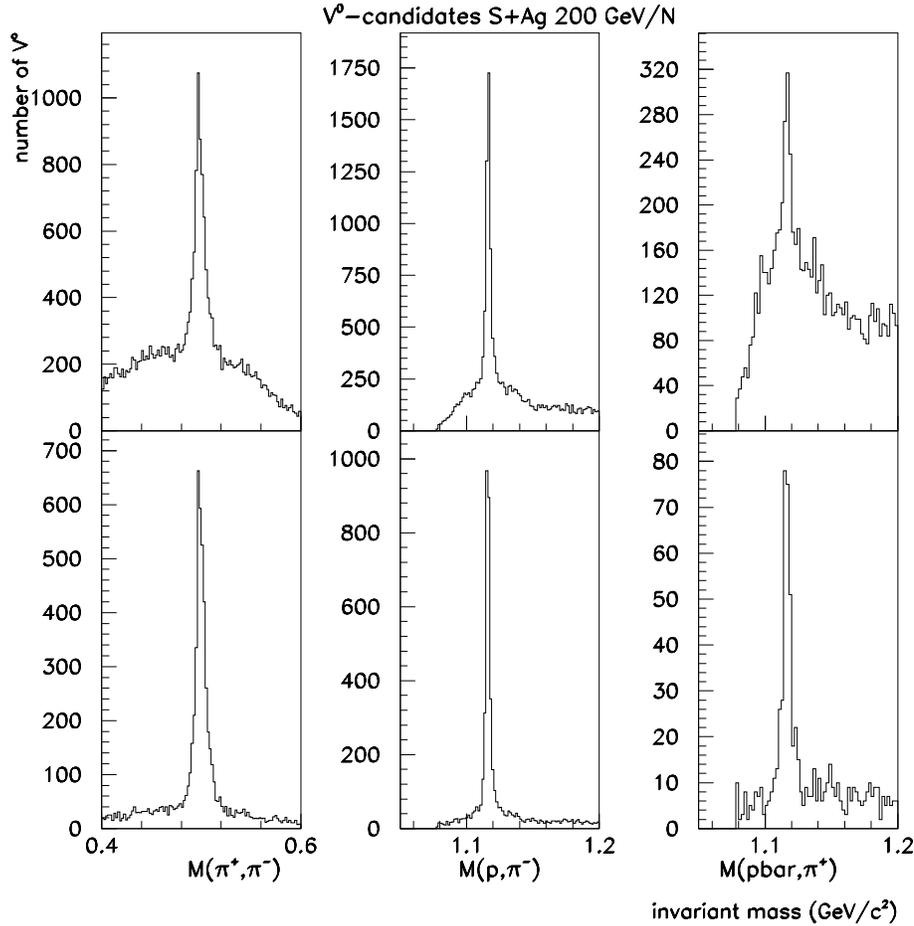,width=12cm}
        \caption[xx]{The invariant mass distributions for
$\Lambda$$~\rightarrow~p~+~\pi^-$,
$\overline{\Lambda}$$~ \rightarrow~\overline{p}~+~\pi^+$  and
$K^0_S$$~\rightarrow~\pi^+~+~\pi^-$  decay
hypothesis for candidates for neutral strange particle
decays (V$^0$ decay) in central
$^{32}$S+Ag  collisions
at 200 GeV per nucleon.
The upper figures show over a large combinatorial background peaks at the invariant
mass of the decay particle. 
The background was reduced significantly by selecting tracks with
low probability for originating from the interaction vertex and
by geometrical cuts (lower figures). 
The final particle identification was done by performing a kinematical fit.
 The distributions are obtained
using the 'automatic' method [4]
of data analysis.}
\end{figure}
\newpage
\begin{figure}
        \epsfig{figure=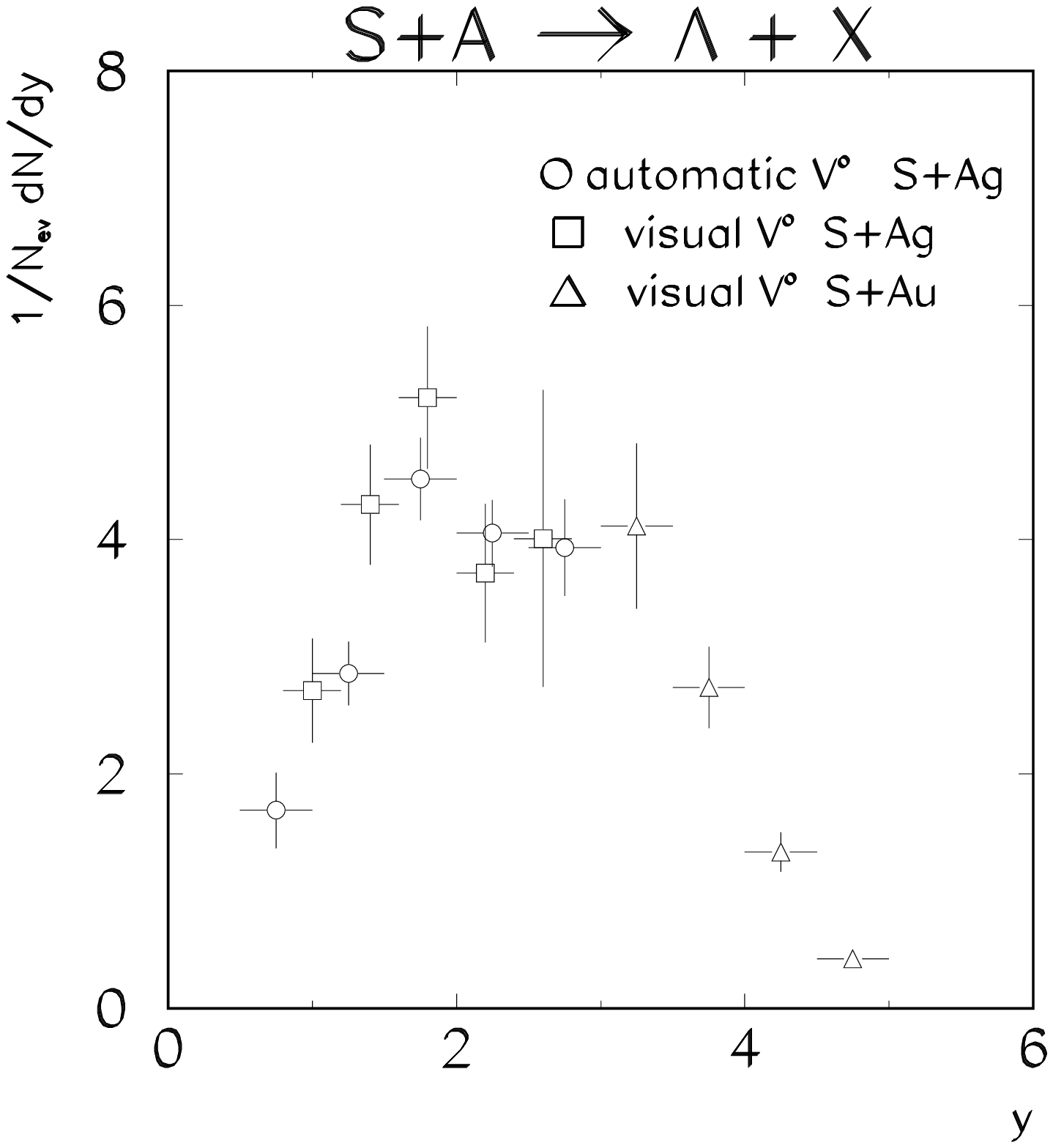,width=9cm,height=5cm}
        \epsfig{figure=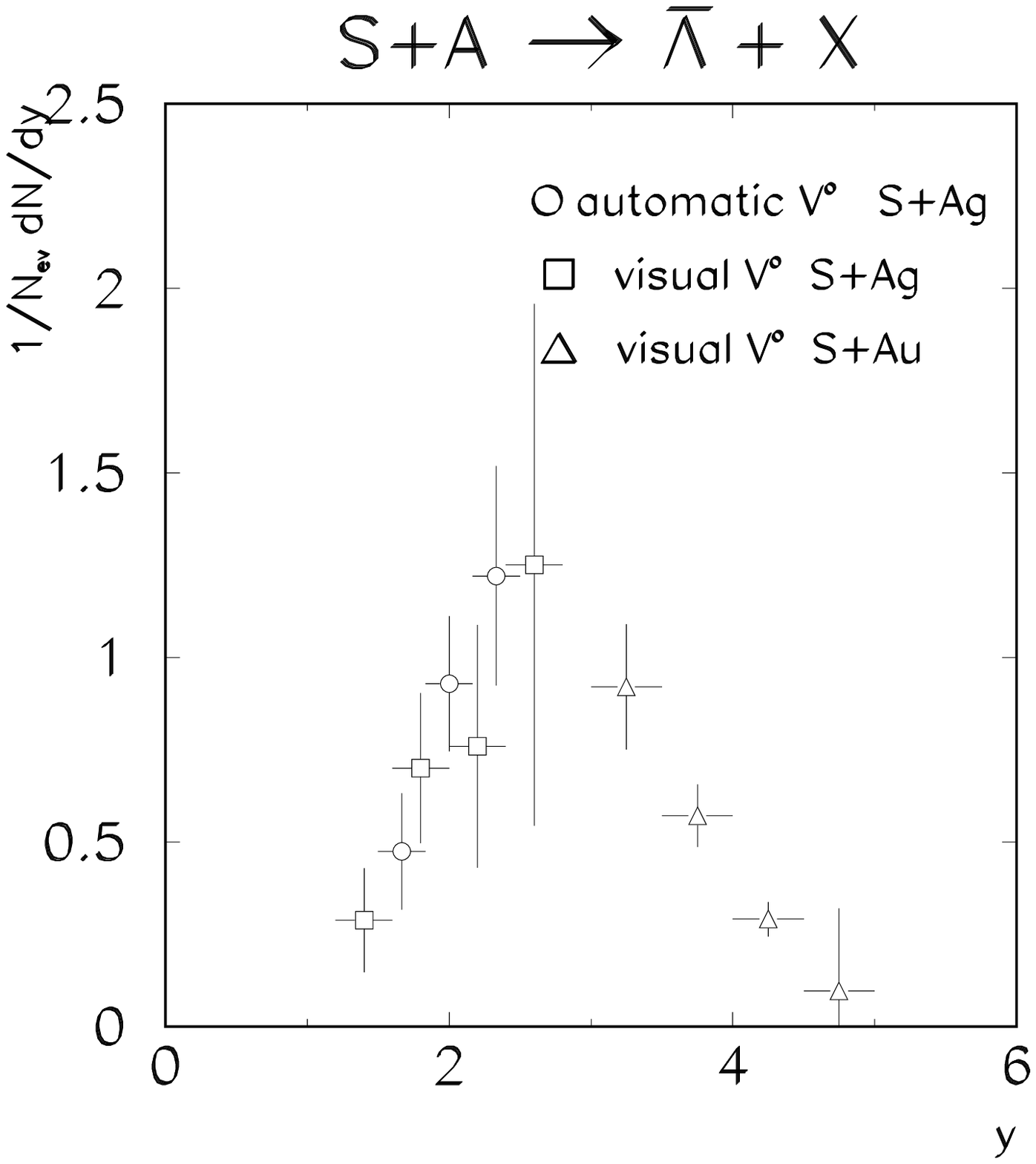,width=9cm,height=5cm}
        \epsfig{figure=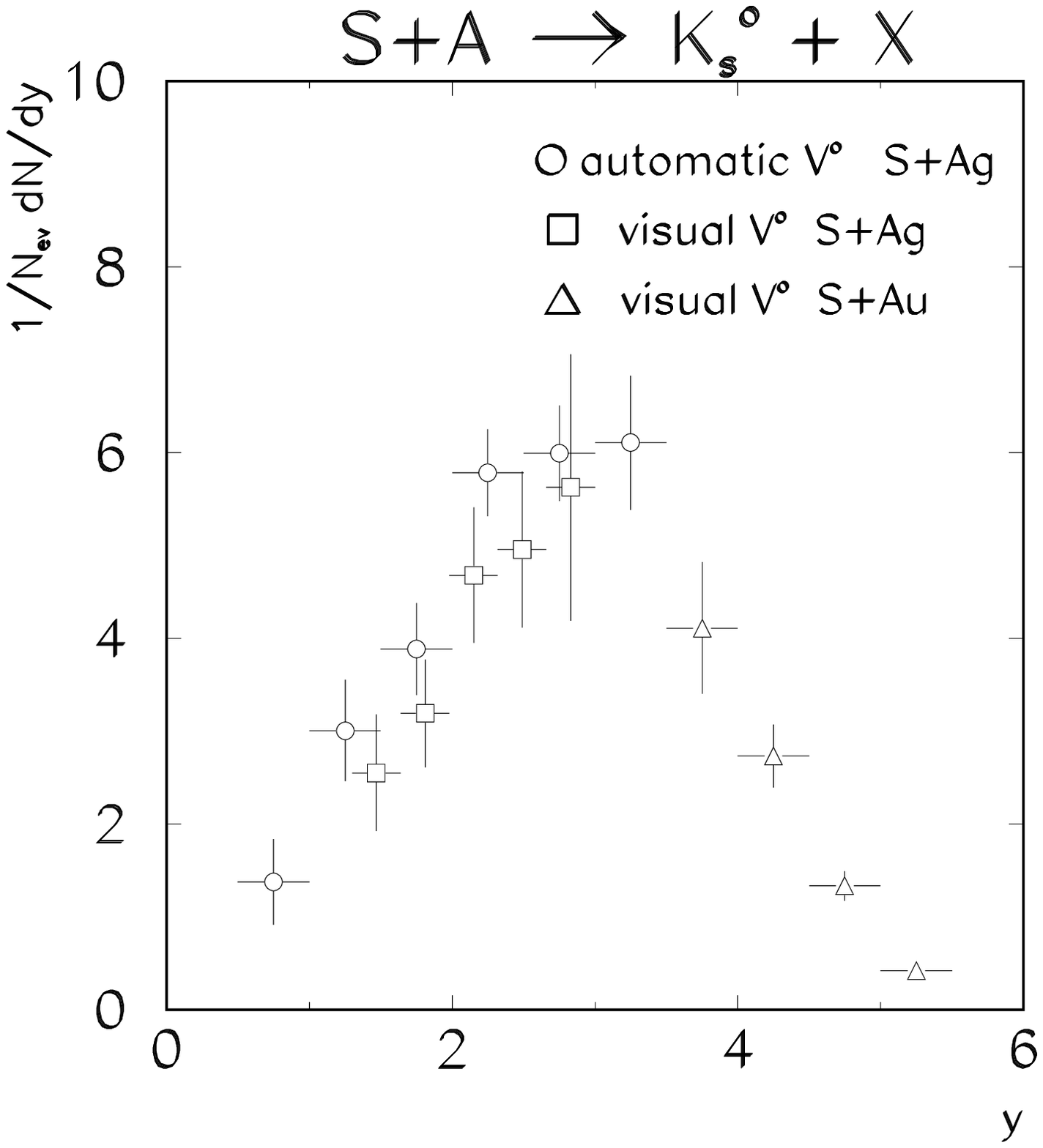,width=9cm,height=5cm}
        \caption[xx]{Rapidity distributions for 
$\Lambda$, $\overline{\Lambda}$ and K$^0_S$.
The figures show the comparison of the 'visual' and 'automatic' 
analysis method for S+Ag collisions in the target hemisphere
(y$<$3) and the results of the 'visual' analysis for S+Au reactions
in the projectile hemisphere (y$>$3) obtained in the 'sweeper'
configuration of the experiment.}
\end{figure}
\newpage

\begin{figure}
        \mbox{
        \epsfig{figure=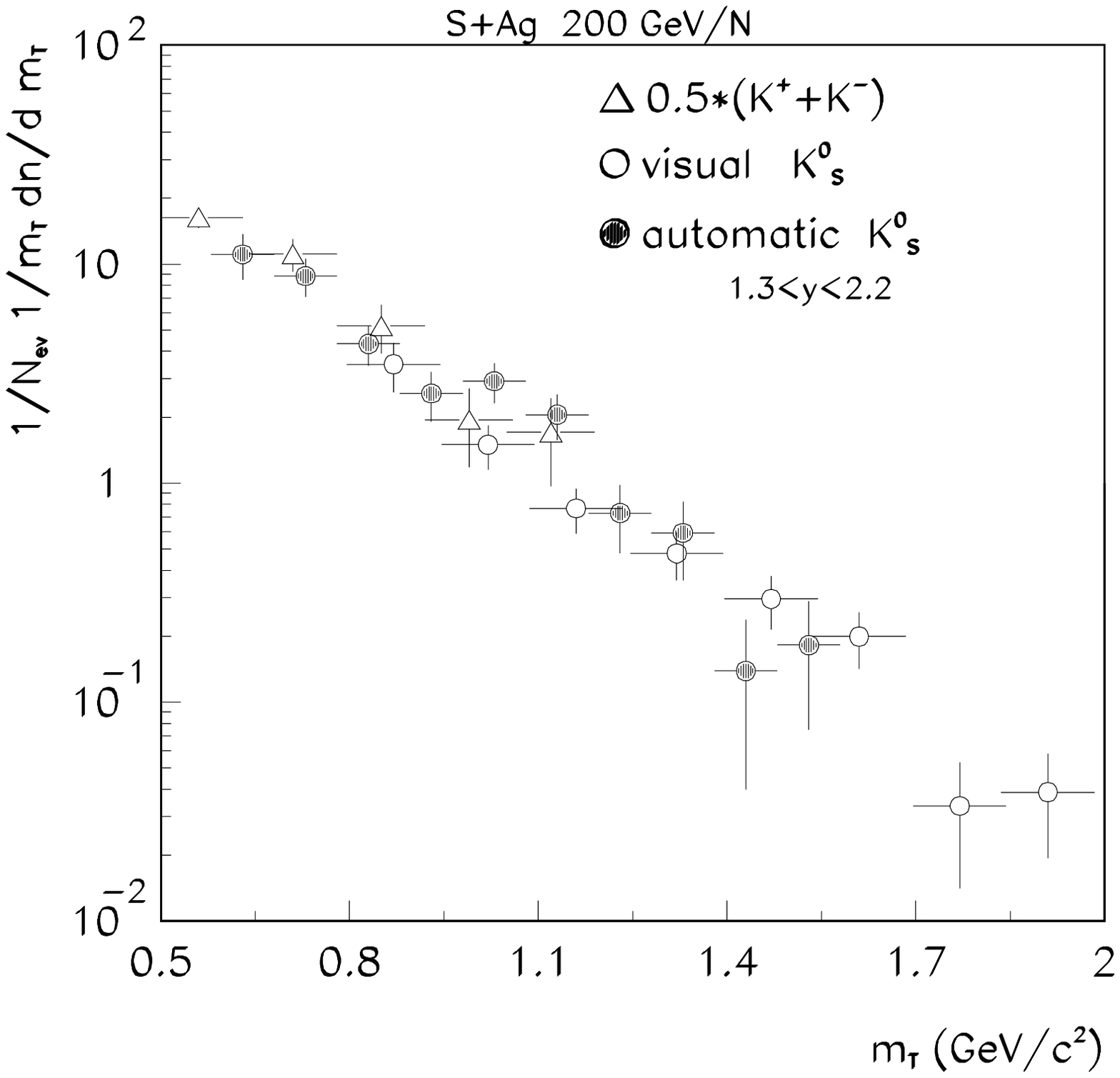,width=6cm}
        \epsfig{figure=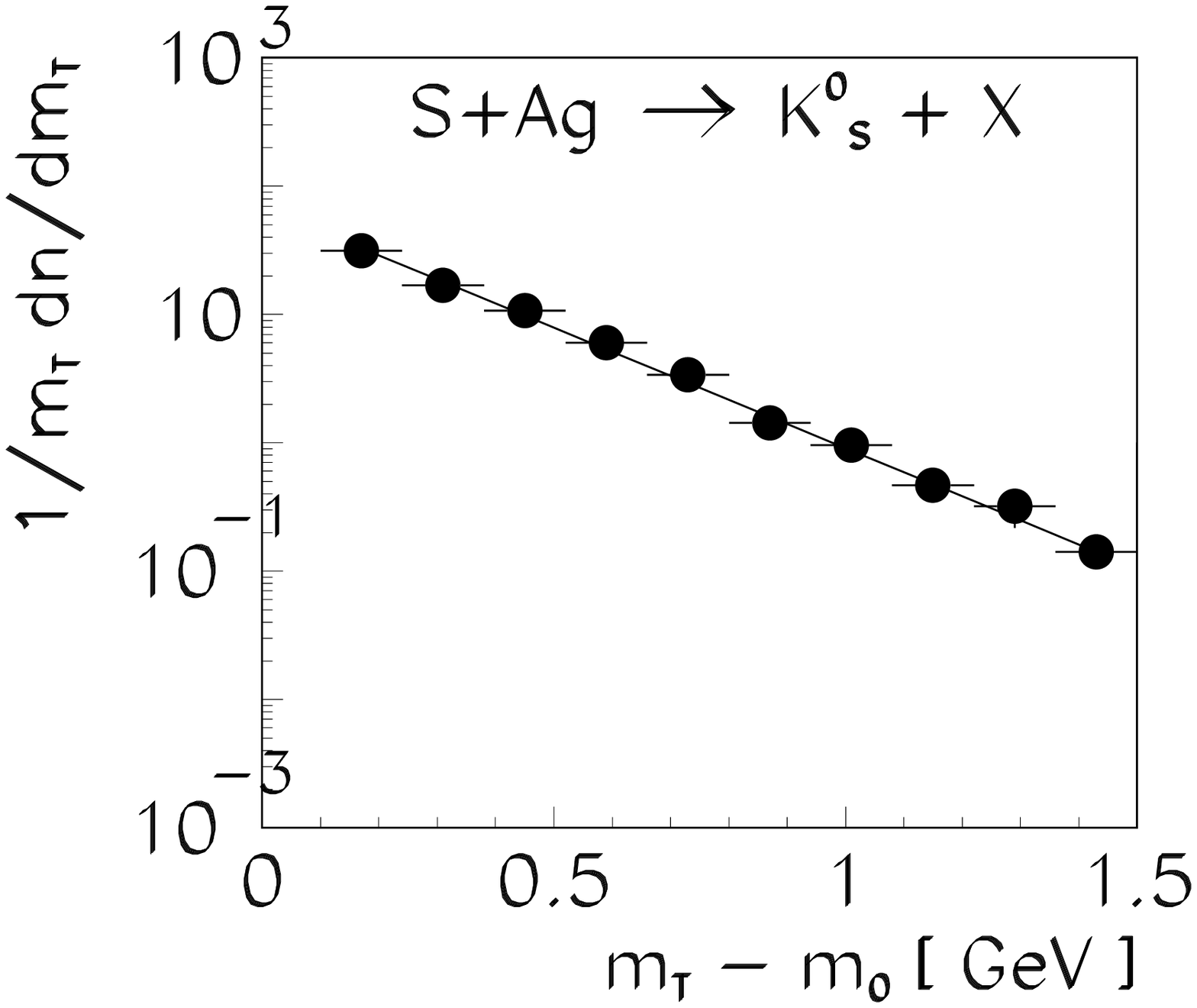,width=6cm,height=6cm}
	}
        \mbox{
        \epsfig{figure=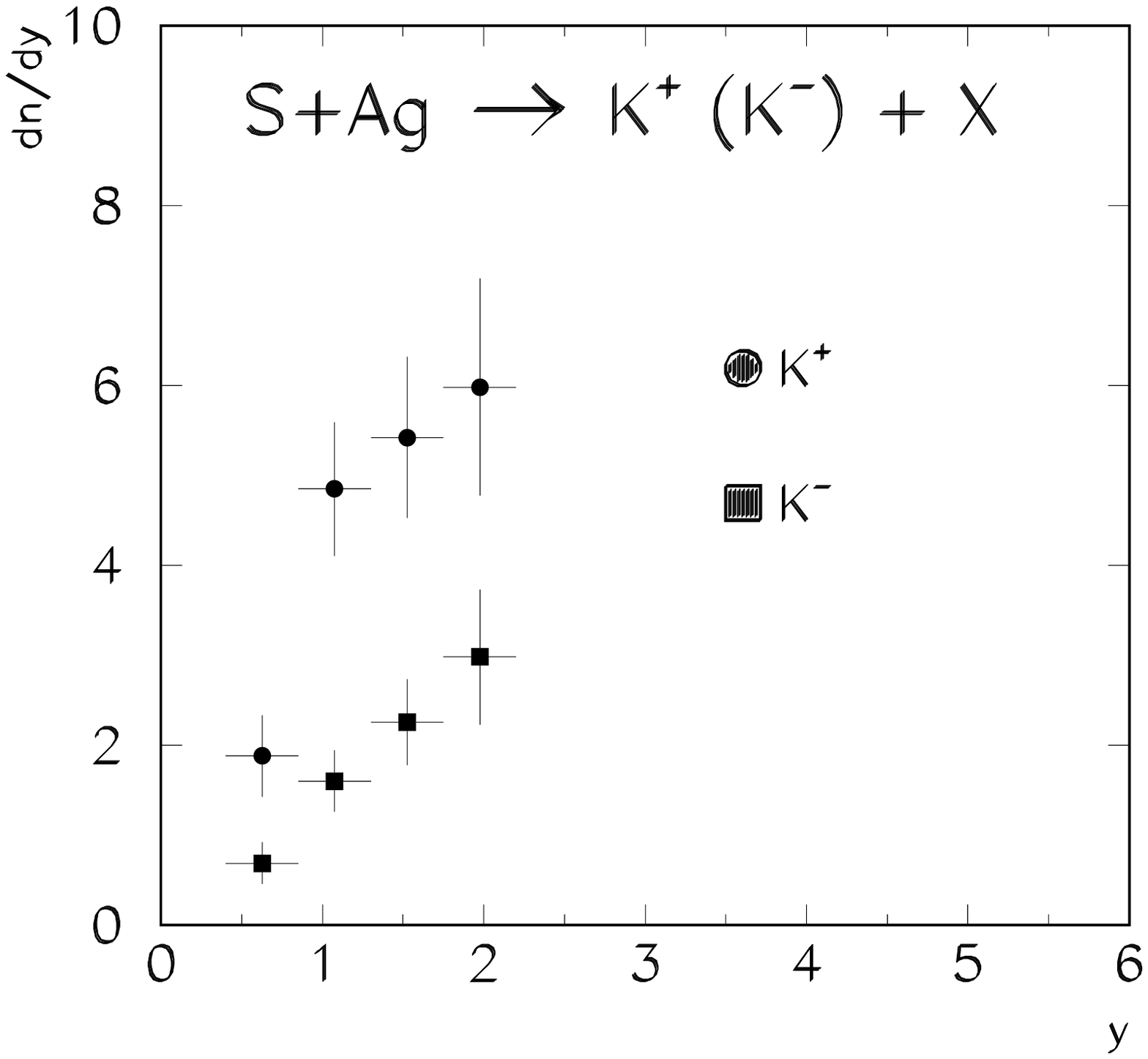,width=6cm}
        \epsfig{figure=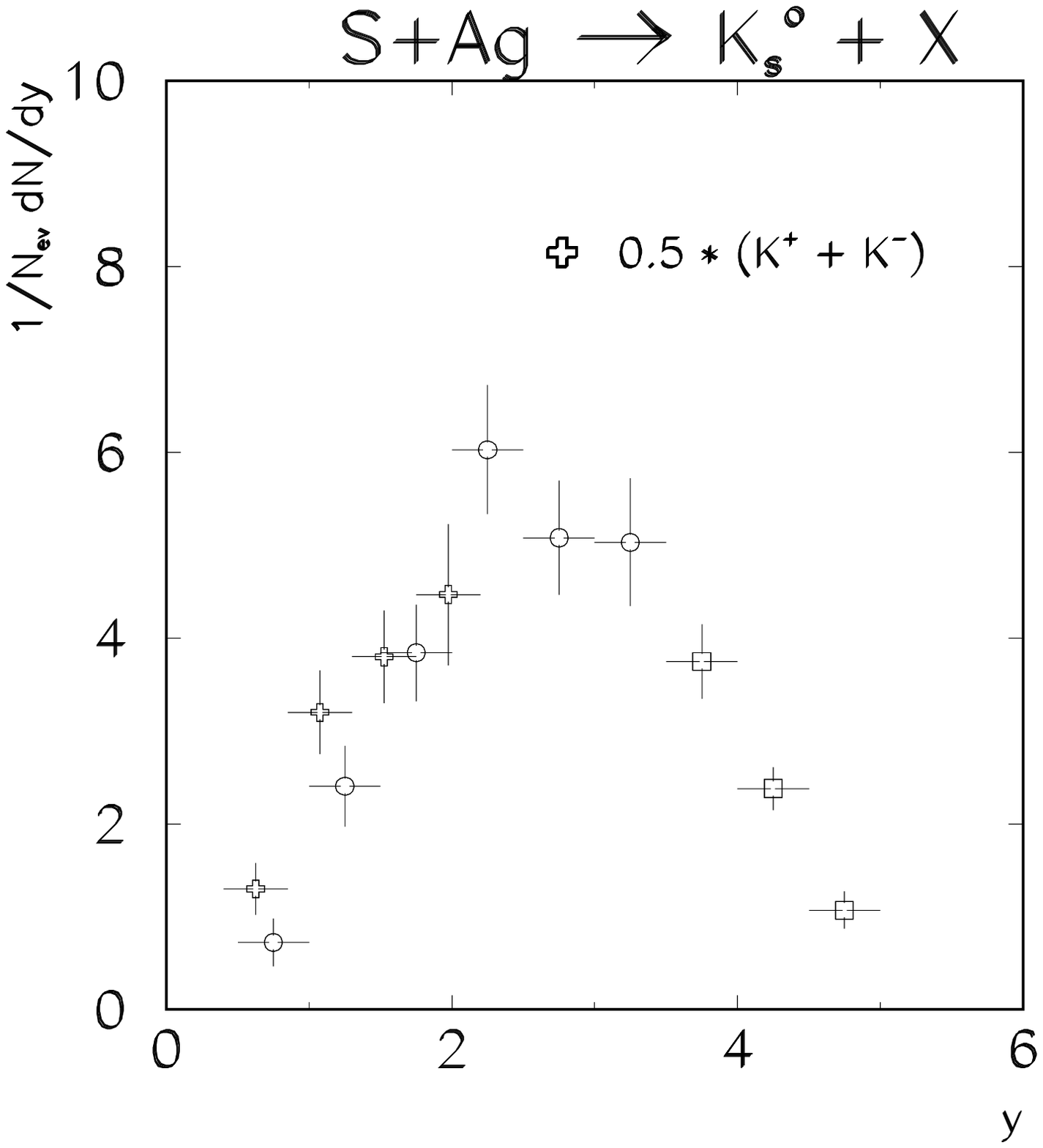,width=6cm,height=5cm}
	}
        \caption[xx]{Kaon systematics:
The upper left figure shows the transverse mass distribution
for kaons in S+Ag collisions in the common rapidity interval. The data for 
(K$^+$+K$^-$)$\cdot$0.5 agree with the 'visual' and 'automatic'
K$^0_S$. This demonstrates the exponential behaviour of the distribution
down to p$_T$=0 (m$_T$=m$_0$).
The m$_t$-distribution of K$^0_S$ (upper right figure) was fitted by an
exponetial parametrization in m$_T$ [2]. The inverese
slope parameter is 231$\pm$17~MeV.
The lower left figure shows the rapidity distribution for charged
kaons [5].
The K$^0_S$ yield has to agree with the average of
K$^+$ and K$^-$  (~0.5$\cdot$(K$^+$+K$^-$)~) for an isospin zero system.   
The lower right figure demonstrates that this is also fullfilled for
the S+Ag system, which has a relatively small charge asymmetry.}
\end{figure}
\newpage
\begin{figure}
        \epsfig{figure=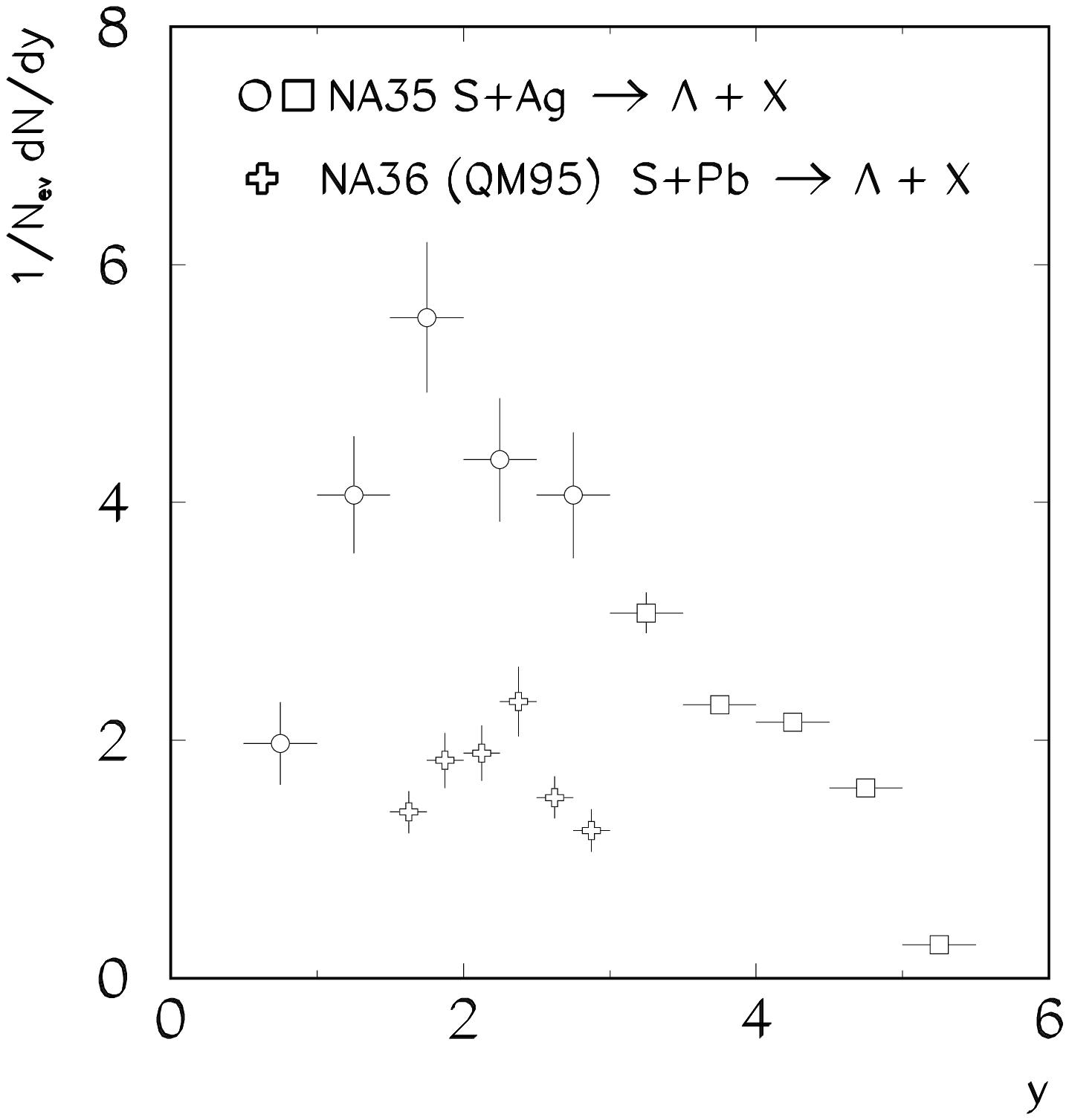,width=9cm,height=4.7cm}
        \epsfig{figure=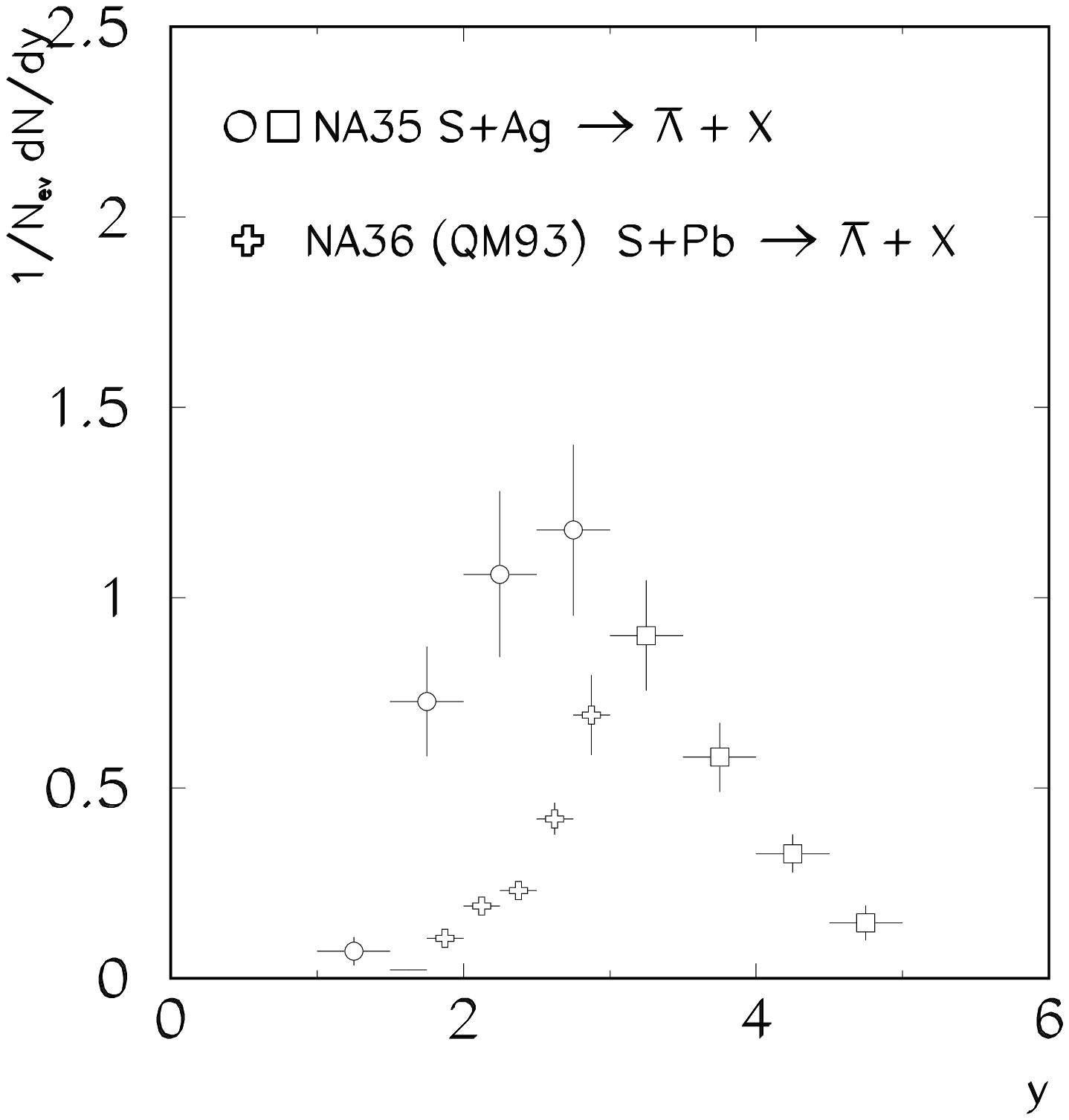,width=9cm,height=4.7cm}
        \epsfig{figure=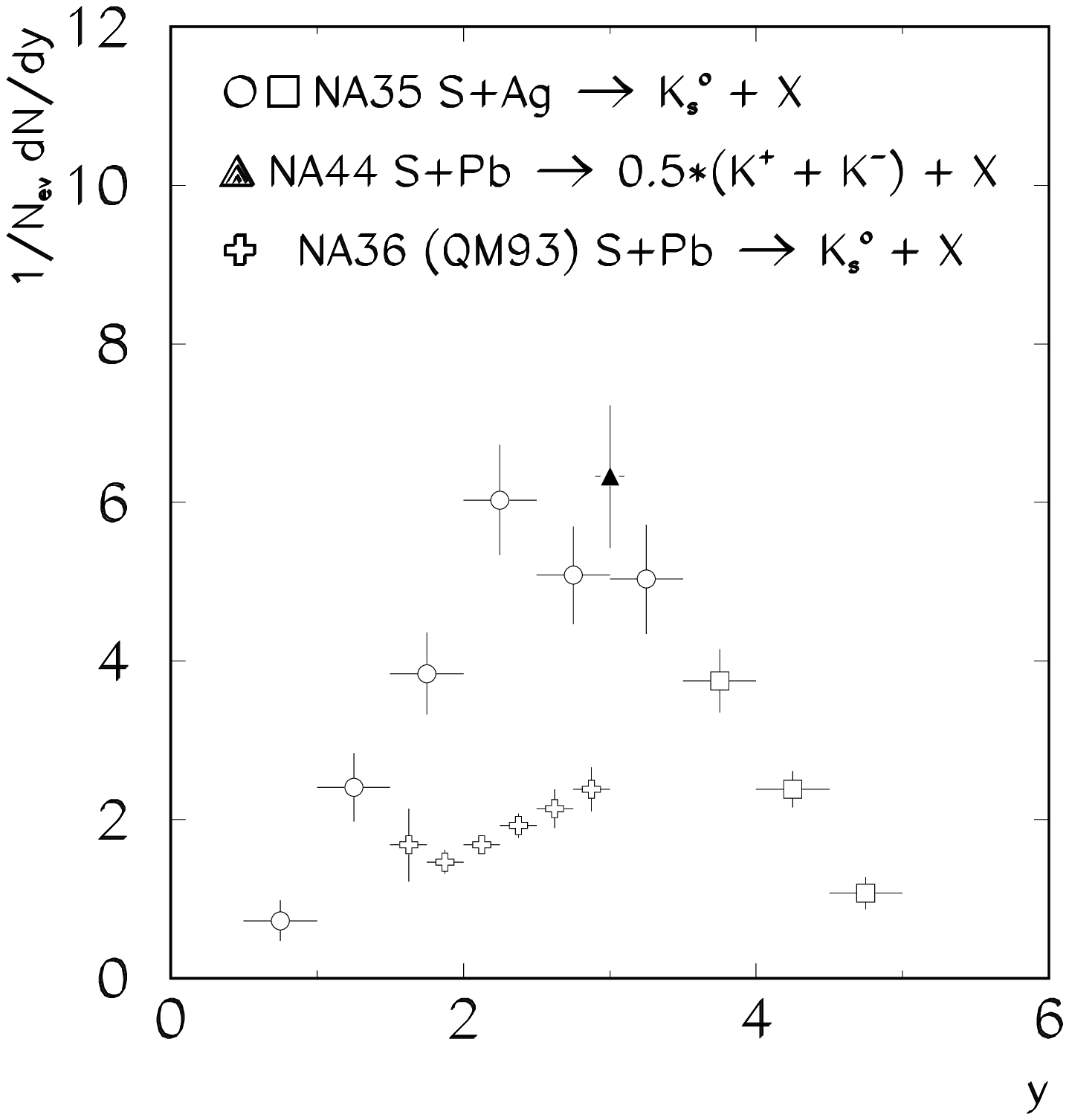,width=9cm,height=4.7cm}
        \caption[xx]{Rapidity distributions for 
$\Lambda$, $\overline{\Lambda}$ and K$^0_S$. 
The NA35 data are compared with the NA36 results in S+Pb collisions.
The NA36 data were scaled in order to correct for the limited
acceptance in transverse momentum. This correction was done
assuming an exponetial behavior im m$_T$. Also a correction for the slightly
softer trigger of NA36 was applied.
The K$^0_S$ data are also compared with the average of
K$^+$ and K$^-$ in S+Pb of NA44 [7],
which confirms the NA35 result,
whereas the NA36 data are more than a factor 2 lower.
}
\end{figure}
\newpage
\begin{figure}
\mbox{
        \epsfig{figure=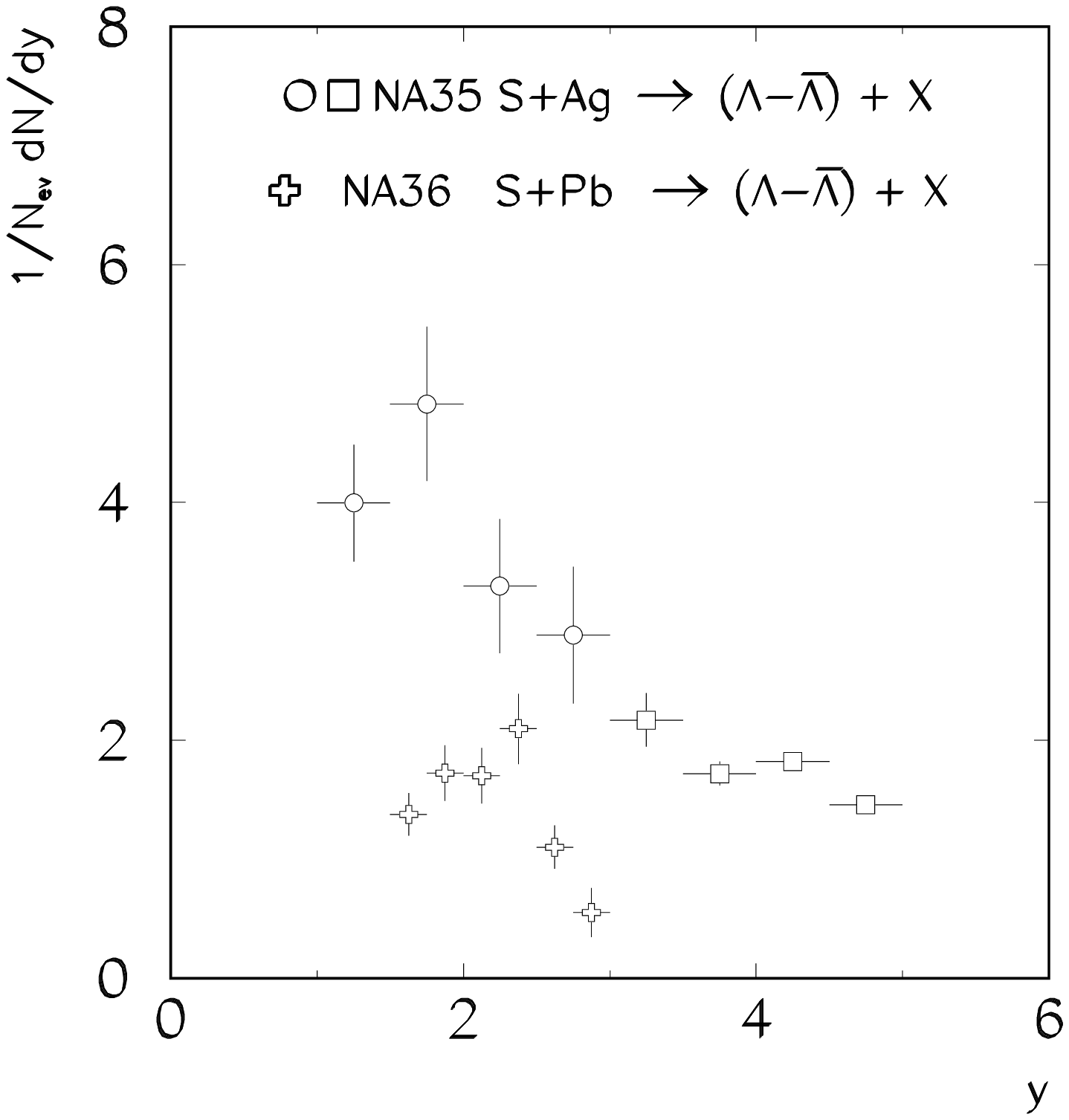,width=6cm,height=5cm}
        \epsfig{figure=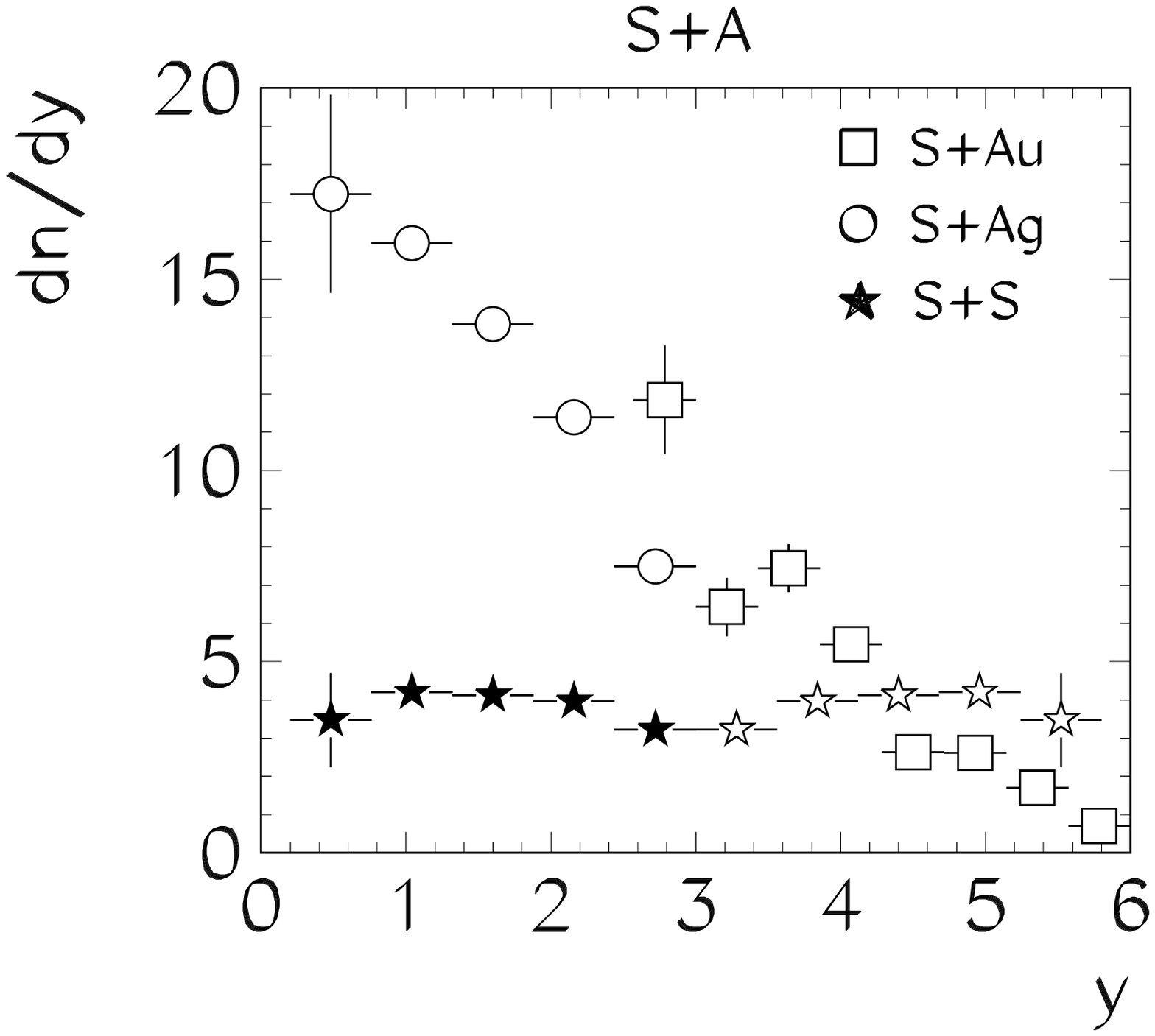,width=6cm,height=5cm}
}
        \caption[xx]{Rapidity distributions for ($\Lambda$-$\overline{\Lambda}$) of 
NA35 and NA36 (left figure). The right figure shows the rapidity distribution
of the 'net protons' ($p-\overline{p}$) for S+S, S+Ag and S+Au [10]. The distribution of
the 'net' $\Lambda$ of NA35 has a similar shape as seen in the 'net protons', because a large
fraction of the momentum of a produced $\Lambda$ is determined by the leading diquark of the proton.
The $\Lambda$-$\overline{\Lambda}$ distribution of NA36 approaches zero at midrapidity, which
is inconsistant with the measured high baryon density.  }
\end{figure}
\begin{figure}
        \epsfig{figure=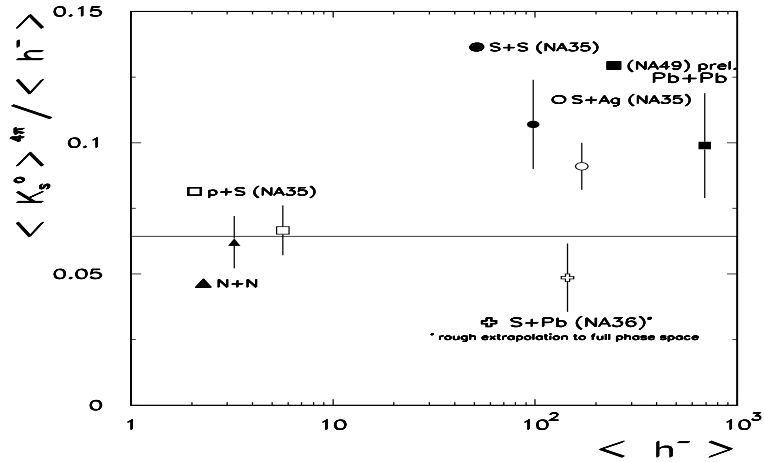,width=10cm,height=6cm}
        \caption[xx]{K$^0_S$ multiplicity in full phase space per negative hadron (h$^-$)
	as a function of h$^-$ for N+N (compilation), p+S, S+S, S+Ag and Pb+Pb.
	The ratio in Nucleus-Nucleus is enhanced by about a factor 1.6 compared to N+N and p+S collisions.
	The NA36 result for S+Pb is below the N+N value. }
\end{figure}
\newpage
\begin{figure}
        \epsfig{figure=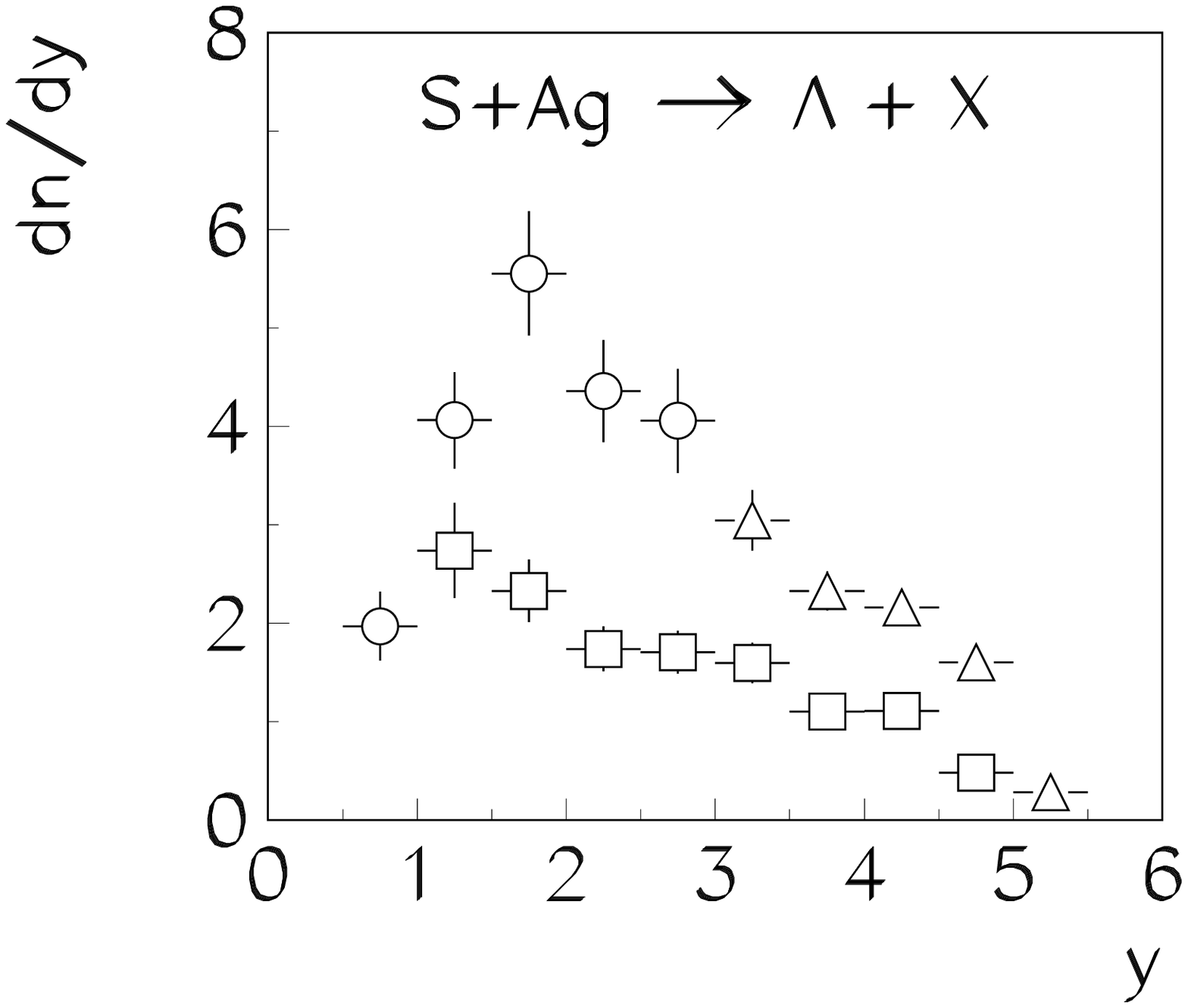,width=9cm,height=5cm}
        \epsfig{figure=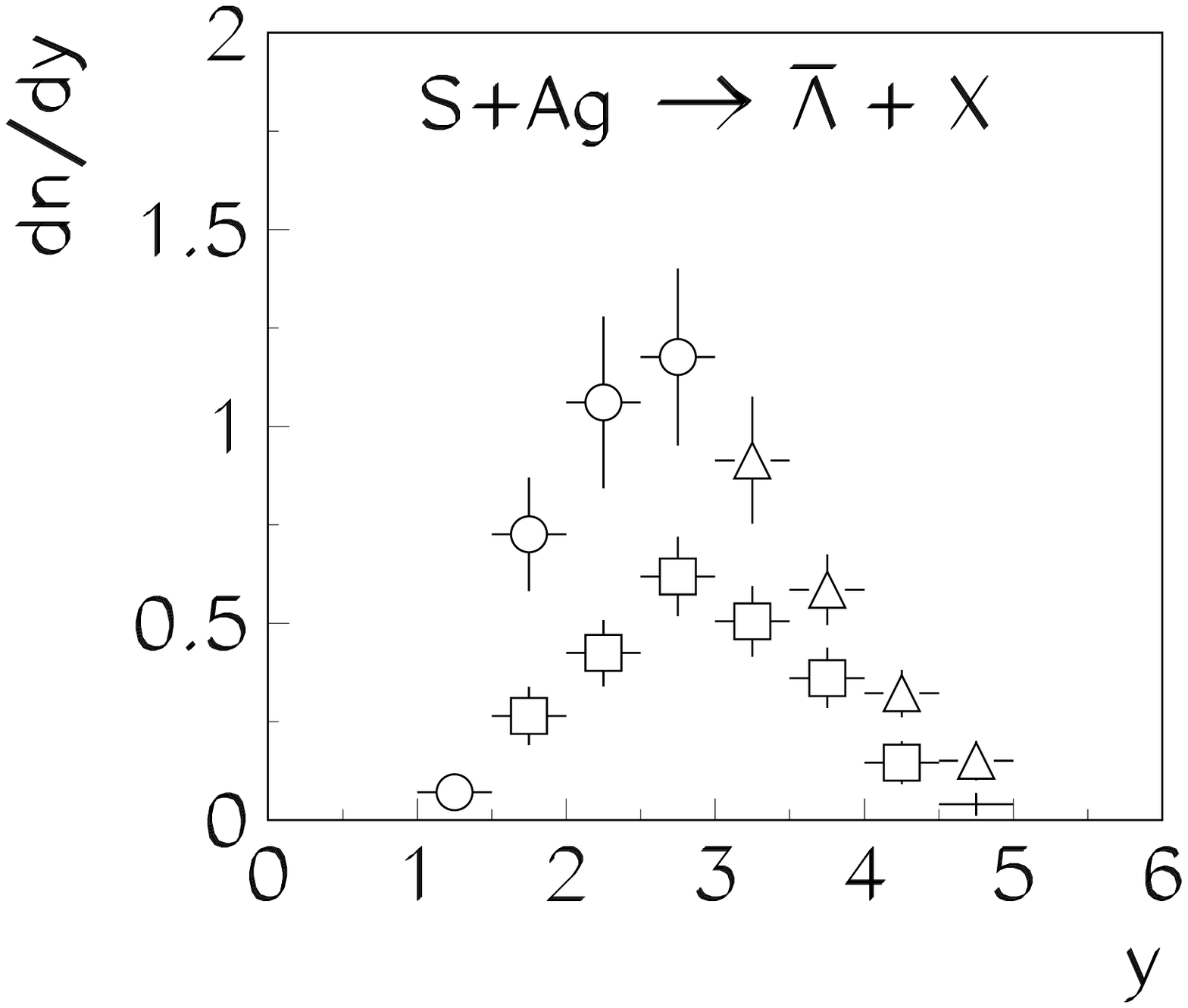,width=9cm,height=5cm}
        \epsfig{figure=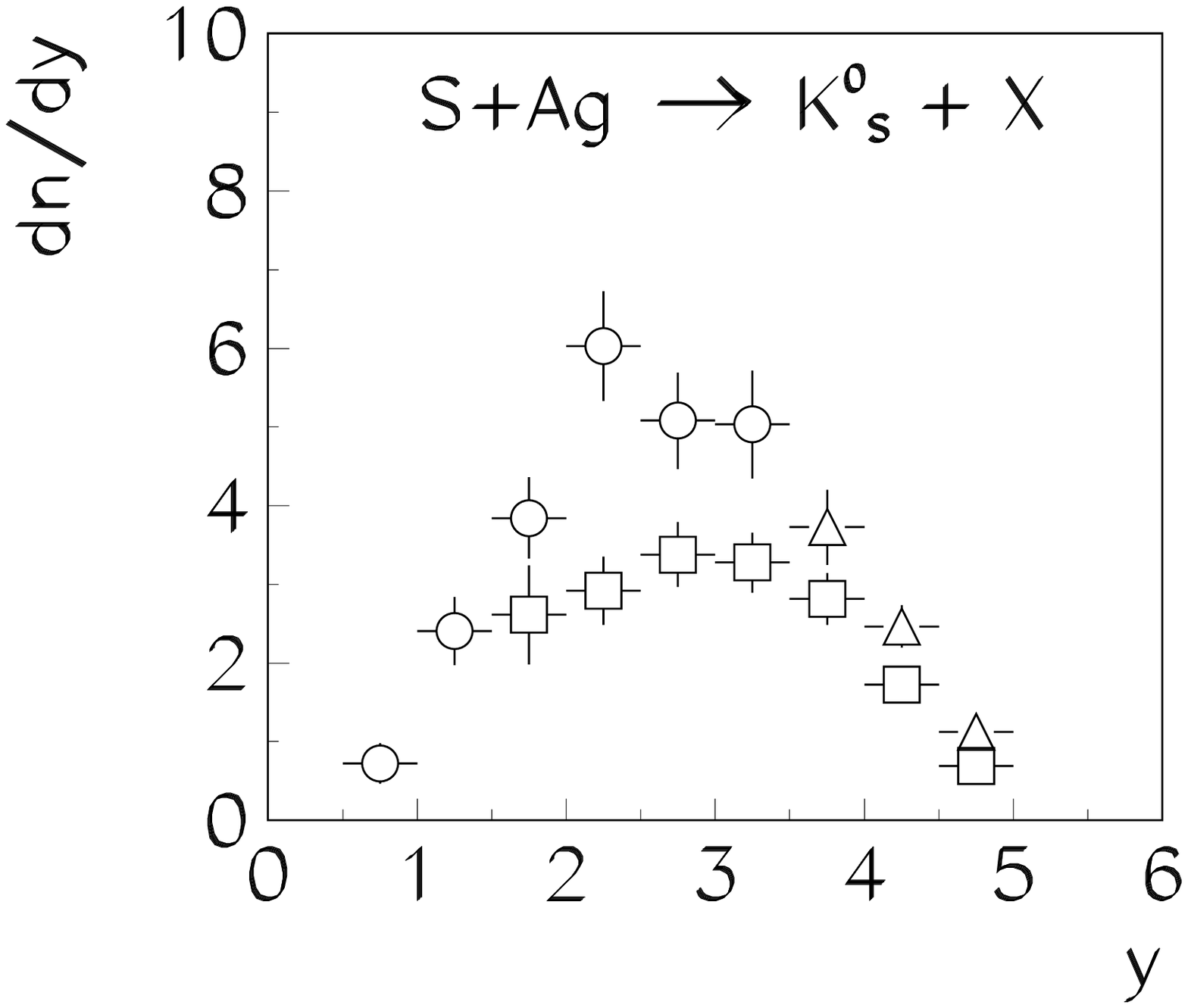,width=9cm,height=5cm}
        \caption[xx]{Rapidity distributions for 
$\Lambda$, $\overline{\Lambda}$ and K$^0_S$ in S+Ag collisions
(open circles and triangles).
	The S+Ag data are compared with particle yields in
	p+S (open squares),
	which were scaled with the ratio of the negative hadron
	multiplicity in the two systems. A clear enhancement of
	about a factor 2 at midrapidity can be seen.}
\end{figure}

\end{document}